\documentclass[a4paper,11pt]{article}
\usepackage{latexsym}
\usepackage{amsmath, amsthm, amssymb, bbm}
\usepackage[mathscr]{eucal}
\usepackage{fancyhdr}
\usepackage{tikz}
\usepackage{cite}
\usepackage{hyperref}
\usepackage{color}
\usepackage[margin=2.85cm]{geometry}

\definecolor{darkred}{rgb}{0.8,0.1,0.1}
\hypersetup{
     colorlinks=false,         
     linkcolor=darkred,
     citecolor=blue,
}

\usepackage{authblk}
\usepackage{slashed}

\usepackage[all,cmtip]{xy}

\theoremstyle{plain}
\newtheorem{proposition}{Proposition}

\theoremstyle{definition}

\theoremstyle{remark}
\newtheorem{remark}[proposition]{Remark}

\newcommand{\rhs}{r.h.s.\ }
\newcommand{\lhs}{l.h.s.\ }
\newcommand{\wrt}{w.r.t.\ }
\newcommand{\cf}{cf.\ }

\newcommand{\ud}{\mathrm{d}}
\newcommand{\del}{\partial}

\DeclareMathOperator{\supp}{supp}

\newcommand{\R}{\mathbb{R}}

\newcommand{\Z}{\mathbb{Z}}

\newcommand{\A}{\mathfrak{A}}

\newcommand{\skal}[2]{\langle #1 , #2 \rangle}

\newcommand{\id}{\mathrm{id}}
\newcommand{\oone}{\mathbbm{1}}
\newcommand{\HS}{\mathcal{H}}

\newcommand{\bra}[1]{\langle #1 |}
\newcommand{\ket}[1]{| #1 \rangle}

\newcommand{\eps}{\varepsilon}

\newcommand{\Spin}{\mathrm{Spin}}

\newcommand{\Dslash}{\slashed{D}}
\newcommand{\Aslash}{\slashed{A}}

\DeclareMathOperator{\tr}{tr}

\newcommand{\ret}{{\mathrm{r}}}
\newcommand{\adv}{{\mathrm{a}}}

\newcommand{\CatAlg}{\mathbf{Alg}}

\newcommand{\CatBG}{\mathbf{Bg}}

\newcommand{\g}{\mathfrak{g}}
\newcommand{\p}{\mathfrak{p}}
\newcommand{\T}{\mathfrak{T}}

\DeclareMathOperator{\Ad}{Ad}
\DeclareMathOperator{\rank}{rank}

\newcommand{\Lie}{\mathcal{L}}

\newcommand{\TO}{\mathcal{T}}
\newcommand{\Ret}{\mathcal{R}}

\DeclareMathOperator{\rce}{rce}
\DeclareMathOperator{\End}{End}

\newcommand{\YM}{{\mathrm{YM}}}

\newcommand{\beq}{\begin{equation}}
\newcommand{\eeq}{\end{equation}}
\newcommand{\nn}{\nonumber}

\begin{document}

\title{Global anomalies on Lorentzian space-times}
\author[1]{Alexander Schenkel\thanks{alexander.schenkel@nottingham.ac.uk}}
\author[2]{Jochen Zahn\thanks{jochen.zahn@itp.uni-leipzig.de}}

\renewcommand\Affilfont{\itshape\small}
\affil[1]{School~of~Mathematical~Sciences,~University~of~Nottingham, 
University~Park,~Nottingham~NG7~2RD,~United~Kingdom}
\affil[2]{Institut~f\"ur~Theoretische~Physik,~Universit\"at~Leipzig,~Br\"uderstr.~16,~04103~Leipzig,~Germany}

\date{March 31, 2017}

\maketitle

\begin{abstract}
\noindent We formulate an algebraic criterion for the presence of global anomalies on globally hyperbolic space-times in the framework of locally covariant field theory. We discuss some consequences and check that it reproduces the well-known global $SU(2)$ anomaly in four space-time dimensions.
\end{abstract}

\section{Introduction and summary}

Global anomalies are an interesting aspect of quantum field theory, 
as they constitute a non-perturbative effect and are thus one of the few 
aspects of this regime which are accessible with our current theoretical tools. 

Global anomalies were first treated in a path integral formalism \cite{Witten82} 
and manifest themselves as the non-invariance of the fermion path integral 
under large gauge transformations (in contrast to the well-known local anomalies,
which occur for infinitesimal gauge transformations). Concretely, chiral fermions 
in four dimensional space-time, charged in the fundamental representation of 
$G = SU(2)$ were considered. As $\pi_4(SU(2)) = \Z_2$, there are compactly 
supported gauge transformations $g$ that can not be deformed to the identity by
compactly supported homotopies. 
But of course one may deform $A$ to its gauge transform $A^g$ via a path $A_\lambda$ of 
connections that are not gauge equivalent to $A$. By studying the flow of 
eigenvalues of the corresponding Dirac operator $\Dslash_{A_\lambda}$ 
along such a deformation, and using the mod 2 index theorem, it was shown 
that the fermion path integral
\[
 \left[ \int \ud \psi \ud \bar \psi \exp( \bar \psi i \Dslash_{A_\lambda} \psi) \right]^{\frac{1}{2}} = \left[ \det i \Dslash_{A_\lambda} \right]^{\frac{1}{2}}
\]
changes sign as $A$ is varied to $A^g$ (note that the path integral for 
chiral fermions is defined as the square root of that for Dirac fermions). 
This implies that the full partition function
\[
 Z = \int \ud A \left[ \det i \Dslash_A \right]^{\frac{1}{2}} \exp \left( - \tfrac{1}{2 g_{\YM}^2} \int \tr F \wedge \star F \right)
\]
vanishes, as the contributions from $A$ and $A^g$ always cancel. 
The theory is thus inconsistent. 

Apart from the ill-definedness of the path integral, the path integral formulation has 
the disadvantage that one considers fermions in non-trivial background fields, 
where the relation between the Euclidean and the Lorentzian setting is unclear.

There is also a Hamiltonian formulation 
\cite{Witten82, JackiwLesHouches, NelsonAlvarezGaume} of global anomalies. 
By choosing temporal gauge, a non-trivial element $g \in \pi_4(G)$ can be transformed 
into a non-trivial element of $\pi_1(\tilde G)$ with $\tilde G = C^\infty_c(\R^3, G)$ 
the group of compactly supported spatial gauge transformations. A global anomaly 
occurs if the implementer of this non-trivial element is not the identity, as there is 
then no gauge invariant state. For the actual computation, one again treats the 
gauge field as a background field. But for time-dependent background fields, 
the Hamiltonian framework is not well suited
as the time-evolution can in general not be implemented on a fixed (vacuum) 
Hilbert space \cite{Ruijsenaars77}.

The aim of the present paper is to give a criterion for global anomalies which 
mimics the above criteria but is properly formulated in a Lorentzian setting. We 
use the framework of locally covariant field theory \cite{BrunettiFredenhagenVerch, HollandsWaldWick} 
which proved very fruitful for quantum field theory in curved space-times or other non-trivial 
backgrounds. Concretely, we use a generalization \cite{LocCovDirac} which also includes 
background gauge fields. Our formulation is based on the implementability of the 
\emph{relative Cauchy evolution} \cite{BrunettiFredenhagenVerch, FewsterVerch12} 
and also relies on the concept of \emph{perturbative agreement} 
\cite{HollandsWaldStress, BackgroundIndependence}.

As for the formulation in the path integral or the Hamiltonian framework, we define and 
compute global anomalies in the setting of free fermions in non-trivial background gauge 
fields. In this setting, a global anomaly does not spoil the consistency of the theory. But as
in the path integral and the Hamiltonian formulation, one expects a global anomaly to spoil 
the consistency of the full theory, involving dynamical gauge fields. We will only briefly 
comment on how this happens in a locally covariant Lorentzian framework.

The article is structured as follows: In the next section, we recall the structures of locally 
(gauge) covariant field theory that are relevant for our discussion. In Section~\ref{sec:GlobalAnomalies} 
we formulate our criterion for global anomalies. In order to make contact with the Hamiltonian 
formulation of global anomalies, we discuss their appearance in Hilbert space representations in 
Section~\ref{sec:Representations}. In particular, we sketch how global anomalies render a gauge 
theory inconsistent. In Section~\ref{sec:1d}, we discuss a toy model where all the constructions 
and the actual computation can be performed explicitly, namely real fermions on the line. To 
prepare for the computation of global anomalies in higher dimensions, we clarify in Section~\ref{sec:WZ} 
the relation between perturbative agreement and the Wess-Zumino consistency condition,
a result that may be of interest independently of the current work. Finally, 
in Section~\ref{sec:Computation}, we compute the $SU(2)$ anomaly in four space-time dimensions, 
using arguments developed in \cite{Witten83a, ElitzurNair}.

\paragraph*{Notations and conventions:}
The symbol $\doteq$ denotes the definition of the \lhs by the r.h.s.. The set 
of smooth compactly supported functions from $M$ to a Lie group $G$ are denoted 
by $C^\infty_c(M, G)$. For $B \to M$ a vector or Lie group bundle over $M$, $\Gamma^\infty_c(M, B)$ 
denotes its set of smooth compactly supported sections. By $\g$ we denote 
the Lie algebra of $G$ and for a principal $G$-bundle $P$ we define $\p$ to be 
the vector bundle associated to the adjoint representation of $G$ on $\g$.
We work in signature $(-, +, \dots, +)$.

\section{The framework}

Let us recapitulate those aspects of the framework of locally covariant field theory 
\cite{BrunettiFredenhagenVerch} which are relevant for our purposes. In contrast
to studying quantum field theories on a fixed background, the basic idea 
of locally covariant field theory is to work coherently over all of them. 
The collection of all possible backgrounds are the objects of the category $\CatBG$; 
its morphisms are used to define relations and consistency conditions
between the theories on different backgrounds.

In the generalized setting of \cite{LocCovDirac}, the admissible backgrounds are tuples 
$(SM, P, \bar g, \bar A)$ consisting of a spin structure $SM$ over an oriented, time-oriented, globally hyperbolic space-time 
$(M, \bar g)$, and a principal $G$-bundle $P$ over $M$ with connection $\bar A$. 
A morphism $\chi: (SM, P, \bar g, \bar A) \to (SM', P', \bar g', \bar A')$ is a tuple 
$(\chi_{SM}, \chi_P)$, where $\chi_{SM}$ is a principal $\Spin_0$-bundle morphism
and $\chi_P$ is a principal $G$-bundle morphism. Both, $\chi_{SM}$ and $\chi_P$ cover the 
same orientation, time-orientation and causality preserving isometric embedding $\psi: M \to M'$. 
Furthermore, the connections are related by pull-back, i.e.\ $\bar A = \chi_P^* \bar A'$. 
In the following, morphisms will often be isomorphisms of the form 
$(\id, \chi_P) : (SM, P, \bar g, \bar A) \to (SM, P', \bar g, \bar A')$,
which affect only the principal $G$-bundle. Typically, these are 
\emph{gauge transformations} $(SM, P, \bar g, \bar A) \to (SM, P, \bar g, \bar A^g)$ 
where $\bar A^g = \Ad_{g^{-1}}  \bar A + g^*(\mu_G)$, with $\mu_G$ the 
Maurer-Cartan form on $G$ and $g \in C^\infty_c(M, G)$.

To each background one assigns an algebra $\A(SM, P, \bar X)$ of observables, 
where we subsumed the geometric data in $\bar X = (\bar g, \bar A)$. The consistency 
of this assignment is encoded in the requirement that $\A : \CatBG \to \CatAlg $ 
is a functor with values in the category $\CatAlg$ of unital $*$-algebras 
(with injective unital $*$-homomorphisms as morphisms). 
This means that for each morphism $\chi : (SM, P, \bar X) \to (SM', P', \bar X')$ 
there is an injective unital $*$-homomorphism
\[
 \alpha_\chi : \A(SM, P, \bar X) \longrightarrow \A(SM', P', \bar X'),
\]
which is an isomorphism if $\chi$ is an isomorphism in $\CatBG$. When $(SM, P)$ 
are kept fixed, also the notation $\A(\bar X)$ will be used.

In order to relate particular observables on different backgrounds, 
one introduces the notion of \emph{fields}, which are natural transformations 
from suitable functors $\T$ to $\A$. Important examples for $\T$ are the functors 
which assign to each background $(SM, P, \bar X)$ the set of test tensors (of a fixed type) on $M$, i.e., 
the set of smooth, compactly supported sections of a vector bundle associated to the bundle $SM + P \to M$. 
A field $\Phi$, restricted to a background $(SM, P, \bar X)$, and smeared with a test tensor $t$, 
is then denoted as $\Phi_{(SM, P, \bar X)}(t)$, or, if $(SM, P)$ are held fixed, also by 
$\Phi_{\bar X}(t)$. Being a natural transformation amounts to
\beq
\label{eq:Field}
 \alpha_\chi \Phi_{(SM, P, \bar X)}(t) = \Phi_{(SM', P', \bar X')}(\chi_* t)
\eeq
for any $t$ and any morphism $\chi: (SM, P, \bar X) \to (SM', P', \bar X')$. A typical example is the \emph{current}
\beq
\label{eq:FermionCurrent}
 j_{(SM, P, \bar X)}(A) \doteq - \int_M \bar \psi \Aslash \psi
\eeq
of free fermions charged under $G$. Here $A \in \Gamma^\infty_c(M, \p \otimes T^* M)$. 
For the explicit construction of the algebra $\A(SM, P, \bar X)$ (in terms of evaluation functionals) 
and fields, in particular non-linear ones, we refer to \cite{LocCovDirac}.

In the setting that we are considering, the background fields provide the hyperbolic (wave or Dirac) 
operator for the dynamical (matter) fields. There is then another way to relate the theories on 
different backgrounds. Let us keep $SM$ and $P$ fixed. For compactly supported perturbations 
$X = (g, A)$ of the geometric data $\bar X = (\bar g, \bar A)$, one defines the retarded 
and advanced M{\o}ller operators
\[
 \tau^{\ret / \adv}_{\bar X + X, \bar X} : \A(\bar X) \longrightarrow \A(\bar X + X),
\]
which are $*$-isomorphisms and act as the identity on observables localized in the 
past/future of $\supp X$, where the two backgrounds are identified via Cauchy surfaces. 
If the algebras $\A$ are constructed as evaluation functionals, one may define the 
M{\o}ller operators by the pullback of the retarded/advanced scattering operator on the solution
spaces.
Alternatively, one may define $\tau^{\ret / \adv}$ abstractly by using 
the time-slice axiom \cite{BrunettiFredenhagenVerch}. 

\begin{remark}
In principle, one could also allow for perturbations $X$ that are not compactly supported, 
but only have a certain fall-off behavior. But this would require restrictions on the growth of the field configurations. 
However, the specific form of natural growth conditions depends both on the background $(SM, P, \bar X)$ 
and on the specific form of the equations of motion operator. In order to work model-independently, 
we stick to perturbations with compact support. 
\end{remark}

For what follows, it is crucial that the M{\o}ller operators compose naturally, i.e.,
\begin{equation}
\label{eq:MollerComposition}
 \tau^{\ret / \adv}_{\bar X + X'', \bar X + X'} \circ \tau^{\ret / \adv}_{\bar X + X', \bar X} = \tau^{\ret / \adv}_{\bar X + X'', \bar X}.
\end{equation}
Furthermore, for $\bar X$ and $\bar X + X$ related by a morphism 
$\chi: (SM, P, \bar X) \to (SM, P, \bar X + X)$ which acts as the identity 
outside of a compact set, we have
\begin{equation}
\label{eq:MollerMorphism}
 \tau^{\ret / \adv}_{\bar X +X, \bar X} = \alpha_\chi.
\end{equation}
For generic $X$, one also has \cite[Prop.~3.7]{FewsterVerch12},
\begin{equation}
\label{eq:MollerMorphismIntertwiner}
 \alpha_\chi \circ \tau^{\ret / \adv}_{\bar X + X, \bar X} = \tau^{\ret / \adv}_{\bar X' + \chi_* X,\bar X'} \circ \alpha_\chi,
\end{equation}
where the morphism on the r.h.s.\ is $\chi :  (SM, P, \bar X) \to (SM, P, \bar X')$
and the one on the l.h.s.\ is $\chi : (SM, P, \bar X + X) \to (SM, P, \bar X' + \chi_* X)$.
For a field $\Phi$, we define the infinitesimal retarded/advanced variation as
\beq
\label{eq:Def_delta}
 \delta^{\ret / \adv}_{\bar X} (X) \Phi(t) \doteq \frac{\ud}{\ud \lambda} \tau^{\ret / \adv}_{\bar X, \bar X + \lambda X} 
 \big(\Phi_{(SM,P,\bar X + \lambda  X)}(t)\big) \big |_{\lambda = 0}.
\eeq

Using the M{\o}ller operators, one can define the \emph{relative Cauchy evolution} \cite{BrunettiFredenhagenVerch} 
as the $*$-automorphism
\beq
\label{eq:Def_rce}
 \rce_{\bar X}(X) \doteq \tau^\ret_{\bar X, \bar X + X} \circ \tau^\adv_{\bar X + X, \bar X}
\eeq
of $\A(\bar X)$. In particular, its derivative
\[
 \dot \rce_{\bar X}(X) \doteq \frac{\ud}{\ud \lambda} \rce_{\bar X}(\lambda X) \big |_{\lambda = 0}
\]
is a derivation. As a consequence of \eqref{eq:MollerMorphism} we have that, 
for $\bar X$ and $\bar X + X$ related by a morphism which acts as the identity 
outside of a compact set,
\begin{equation}
\label{eq:rceMorphism}
 \rce_{\bar X}(X) = \id.
\end{equation}

\section{Global anomalies}
\label{sec:GlobalAnomalies}

In the following, we will focus on perturbations of the connection $\bar A$. 
For a free theory in the presence of background fields, the infinitesimal version of the 
relative Cauchy evolution is given by the commutator with the current, 
\beq
\label{eq:CurrentImplementsRceDot}
 \dot \rce_{\bar A}(A) = - \frac{i}{\hbar} [j_{\bar A}(A), \,\cdot\, ].
\eeq
More generally, one may say that the theory $\A:\CatBG \to \CatAlg$ \emph{admits a current} 
if there is a field $j$ such that the above holds.
As a consequence of \eqref{eq:rceMorphism}, $\dot \rce$ vanishes for an infinitesimal 
gauge transformation, i.e.,
\beq
\label{eq:dot_rce_trivial}
\dot \rce_{\bar A}( \bar \ud c) = 0,
\eeq
where $\bar \ud$ is the covariant differential defined on sections of $\p$ by
\[
 (\bar \ud c)(\xi) \doteq \bar \nabla_\xi c,
\]
with $\xi$ a vector field and $\bar \nabla$ the covariant derivative induced 
on the associated bundle $\p$ by the connection $\bar A$. The condition 
that the current $j$ is conserved can then be formulated as the requirement
\beq
\label{eq:CurrentConservation}
 \bar \delta j_{\bar A}(c) \doteq j_{\bar A}(\bar \ud c) = 0.
\eeq
The existence of a conserved current signifies the absence of infinitesimal 
(usually called local) anomalies and is assumed from now on. It is important to note that 
an anomaly can not be seen in a failure of \eqref{eq:dot_rce_trivial}, as a violation of 
\eqref{eq:CurrentConservation} is a c-number.

In order to discuss global anomalies, we
require that the $*$-automorphism 
$\rce_{\bar A}(A)$ is unitarily implemented by a field $V_{\bar A} (A)$, i.e.,
\beq
\label{eq:Def_rce_inner}
 \rce_{\bar A}(A)(a)  = \mathrm{Ad}_{V_{\bar A}(A)}(a) \doteq V_{\bar A}(A)\, a\,  V_{\bar A}(A)^* ,
\eeq
for all $a\in \A(SM,P,\bar g ,\bar A)$, and
\[
V_{\bar A}(A)^{-1}  = V_{\bar A}(A)^*.
\]
The implementer $V_{\bar A}(A)$ can be seen as the inverse of the $S$ matrix 
for scattering at the potential $A$, \cf also the discussion below.
In the present work, we are focusing on conceptual and structural aspects, not on functional analytic ones, 
so we allow the implementers to be formal power series in the perturbation $A$. 
Whether this leads to a proper unitary representation in suitable Hilbert space 
representations is a very interesting open question.\footnote{This is related to the 
question whether the current $j_{\bar A}(A)$ is represented by an essentially
self-adjoint operator on $\HS$. For the scalar field and Wick squares without derivatives, 
conditions ensuring this property were given in \cite{SandersSA}.} The requirement 
that the rce should be unitarily implemented in representations seems to have first been considered 
in \cite{FewsterGlobalGaugeGroup}.

Obviously, if $\rce_{\bar A}$ is implemented 
by $V_{\bar A}$, then there is a current
\beq
\label{eq:Current}
 - \frac{i}{\hbar} j_{\bar A}(A) \doteq \frac{\ud}{\ud \lambda} V_{\bar A}(\lambda A) \big|_{\lambda = 0}.
\eeq
However, as discussed above, one should impose further constraints on 
the current than just \eqref{eq:CurrentImplementsRceDot}.
Hence, apart from the implementers being fields, we impose a further 
natural condition.
To motivate it, we note that
from \eqref{eq:MollerComposition} and \eqref{eq:Def_rce} it follows that
\[
 \rce_{\bar A}(A') = \tau^\ret_{\bar A, \bar A + A} \circ \rce_{\bar A + A}(A' - A) \circ \tau^\adv_{\bar A + A, \bar A}.
\]
Using \eqref{eq:Def_rce_inner}, it follows that
\[
 \rce_{\bar A}(A') = \Ad_{V_{\bar A} (A^\prime)} = 
 \Ad_{ \tau^\ret_{\bar A, \bar A + A}(V_{\bar A + A} (A'-A))\, V_{\bar A}(A)}.
\]
It thus seems natural to require that
\begin{equation}
\label{eq:Vcondition}
 V_{\bar A}(A') = \tau^\ret_{\bar A, \bar A + A}(V_{\bar A + A}(A'-A)) \,V_{\bar A}(A).
\end{equation}
For $A' = A_1 + A_2$, with $\supp A_2$ not intersecting the past of $\supp A_1$, 
this reduces to the causality requirement \cite{BogoliubovShirkov, Scharf95}
\beq
\label{eq:CausalSMatrix}
 S(A_1 + A_2)^{-1} = S(A_1)^{-1} S(A_2)^{-1}
\eeq
for the $S$ matrix, given the identification of $V$ with $S^{-1}$ and the fact that 
$\tau^\ret_{\bar A, \bar A + A_2}$ acts trivially on fields whose support does not 
intersect the future of $\supp A_2$. Note, however, that our condition \eqref{eq:Vcondition} 
is stronger than \eqref{eq:CausalSMatrix} in that no conditions on the relations of 
the supports are made, \cf also the discussion in Remark~\ref{rem:CausalSMatrix} below.
The discussion of Hilbert space representations in the following section 
provides further motivation for the requirement \eqref{eq:Vcondition}. If condition \eqref{eq:Vcondition} 
is fulfilled, we say that $\rce$ is \emph{unitarily implemented} by $V$. 

It turns out that \eqref{eq:Vcondition} is automatically fulfilled for free fields charged 
under $G$ in space-time dimension $d \leq 4$ if 1.) the current 
\eqref{eq:Current} is conserved and 2.) $V_{\bar A}(0) =\oone$ is the unit. 
To see this, note that \eqref{eq:Vcondition} can be equivalently written as
\beq
\label{eq:V_Consistency}
 \tau^\ret_{\bar A, \bar A + A}(V_{\bar A + A}(A''-A)) \, V_{\bar A}(A) = 
 \tau^\ret_{\bar A, \bar A + A'}(V_{\bar A + A'}(A''-A'))\,  V_{\bar A}(A').
\eeq
As the space of connections is affine, there are no topological obstructions to fulfill this condition, 
but there may be local ones. Replacing $A$ by $\lambda A$, $A'$ by $\eta A'$, and $A''$ by $\lambda A + \eta A'$ 
in the above equation and evaluating the derivative \wrt $\lambda$ and $\eta$ at $0$, we obtain
\begin{equation}
\label{eq:E=0}
 E_{\bar A}(A, A') \doteq \delta^\ret_{\bar A}(A) j(A') - \delta^\ret_{\bar A}(A') j(A) - i \hbar^{-1} [j_{\bar A}(A'), j_{\bar A}(A)] = 0,
\end{equation}
where we used the definition \eqref{eq:Def_delta}. As will become clear in the following section, we may see $E_{\bar A}$ 
as the curvature for ``parallel transport'' by ${\bar \pi}(V_{\bar A})$ of vectors $\Psi \in \bar \HS$ in any representation 
${\bar \pi}: \A(\bar A) \to \End(\bar \HS)$.
The condition \eqref{eq:E=0} is known the context of background independence. Its fulfillment 
(possibly after a finite renormalization of the current $j$) is a necessary and sufficient condition 
to fulfill perturbative agreement by inductively redefining time ordered products by finite 
renormalizations \cite{HollandsWaldStress, BackgroundIndependence}.
As shown in \cite[Prop.~3.8]{BackgroundIndependence}, \eqref{eq:E=0} is fulfilled 
for $d \leq 4$, whenever the current $j$ is conserved.

Obvious consequences of \eqref{eq:Vcondition} are
\begin{align}
 \label{eq:V_0}
 V_{\bar A}(0) & = \oone, \\
\label{eq:delV}
 \frac{\ud}{\ud \lambda} V_{\bar A}(A+\lambda A') \big |_{\lambda = 0} & = - \frac{i}{\hbar} \,\tau^\ret_{\bar A, \bar A + A}(j_{\bar A + A}(A')) \,V_{\bar A}(A).
\end{align}
Hence, in order to evaluate $V_{\bar A}(A)$, we may choose any path 
$A_\lambda$ such that $A_0 = 0$, $A_1 = A$ and compute the path-ordered exponential integral
\beq
\label{eq:POIntegral}
 V_{\bar A}(A) = P \exp \left( - \frac{i}{\hbar} \int_0^1 \tau^\ret_{\bar A, \bar A + A_\lambda}(j_{\bar A+ A_\lambda}(\dot A_\lambda)) \, \ud \lambda \right).
\eeq
In particular, for any $\bar A + A$ which is continuously connected by a path of compactly 
supported gauge transformations to $\bar A$, we have that $V_{\bar A}(A) = \oone$. 
Assuming that the center of $\A(SM,P,\bar g,\bar A)$ is trivial, we also know that $V_{\bar A}(A) = e^{i \phi}\, \oone$ 
for some $\phi \in \R$ whenever $\bar A^g = \bar A + A$ for a compactly 
supported gauge transformation $g$.  We say that the theory $\A$ has a 
\emph{global anomaly}, if $V_{\bar A}(A) \neq \oone$ in that case.

For four-dimensional space-times with trivial topology $\mathbb{R}^4$, possible 
obstructions are thus classified by $\pi_4(G)$. Given a non-trivial gauge 
transformation $\bar A \mapsto \bar A^g = \bar A + A$, \eqref{eq:POIntegral} can in principle 
be used to decide whether a given theory is anomalous or not.

Apart from leading to the same obstructions as in the path integral framework, 
our formulation has the additional similarity that the computation of the anomaly 
proceeds by reaching the non-trivial gauge transformation via a path of 
gauge non-equivalent backgrounds.

\begin{remark}
\label{rem:CausalSMatrix}
A conserved current $j$ has the residual renormalization ambiguity 
$j_{\bar A} \to j_{\bar A} + c \bar \delta \bar F$ of adding multiples of the 
current of the background connection ($\bar F$ being its curvature and $\bar \delta$ 
the corresponding covariant divergence) \cite{LocCovDirac}. This corresponds 
to a charge renormalization. Such a transformation obviously changes the implementers as
\beq
\label{eq:Vambiguity}
 V_{\bar A}(A) \to \exp \left( - \frac{i c}{2 \hbar} \int_M \tr \left( F[\bar A + A] \wedge \star F[\bar A + A] - \bar F \wedge \star \bar F \right) \right) V_{\bar A}(A).
\eeq
It follows that the implementers as such are not unique, only for a given renormalization 
prescription of the current. It may be worth pointing out that the causality requirement \eqref{eq:CausalSMatrix}
is much less restrictive: If $S(A)$ is an $S$ matrix fulfilling \eqref{eq:CausalSMatrix}, then
\[
 S(A) \to e^{ i G_{\bar A, \bar g}[A]  } S(A)
\]
yields another one, for any local functional $A \mapsto G_{\bar A, \bar g}[A] \in \R$. 
The point is that \eqref{eq:CausalSMatrix} can not restrict the local functional $G$, as 
the supports of $A_1$ and $A_2$ do never intersect. By contrast, our condition 
\eqref{eq:Vcondition}, combined with the requirement that the current defined by 
\eqref{eq:Current} is conserved, implies severe restrictions on $G$, leading to 
essentially unique implementers, up the the ambiguity \eqref{eq:Vambiguity}.\footnote{In
\cite{ScharfWreszinski}, \cf also \cite[Chapter~2]{Scharf95}, the $S$ matrix is fixed by an explicit
non-perturbative construction, exploiting the causality \eqref{eq:CausalSMatrix} 
by cutting $A$ into temporal slices and performing limits in which first the size of the 
separation and then the size of the slices tends to 0. However, it is not shown that the result
 is independent on how this slicing is performed. It is also not clear whether the thus constructed
 $S$ matrices fulfill our stronger requirements.
 In \cite{GraciaBondia}, it is claimed that the $S$ matrix is unique, even without assuming 
\eqref{eq:CausalSMatrix}. However, the claim is based on the assumption that 
the time evolution $U(s,t)$ in the interaction picture is implementable and the 
implementation strongly differentiable, requirements that seem difficult to justify.}
 \end{remark}

Using \eqref{eq:POIntegral}, we can show that $V$ is indeed a field. Let $\chi : (SM,P,\bar X) \to (SM,P,\bar X')$
be a morphism. We compute
\begin{align}
 \alpha_\chi V_{\bar A}(A) & = P \exp \left( - \frac{i}{\hbar} \int_0^1 \alpha_\chi \circ \tau^\ret_{\bar A, \bar A + A_\lambda}(j_{\bar A+ A_\lambda}(\dot A_\lambda)) \, \ud \lambda \right) \nn \\
 & = P \exp \left( - \frac{i}{\hbar} \int_0^1 \tau^\ret_{\bar A', \bar A' + \chi_* A_\lambda} \circ \alpha_\chi (j_{\bar A+ A_\lambda}(\dot A_\lambda)) \, \ud \lambda \right) \nn \\
 & = P \exp \left( - \frac{i}{\hbar} \int_0^1 \tau^\ret_{\bar A', \bar A' + \chi_* A_\lambda} (j_{ \bar A' + \chi_* A_\lambda}(\chi_* \dot A_\lambda)) \, \ud \lambda \right) \nn \\
\label{eq:V_Field}
 & = V_{\bar A'}(\chi_* A).
\end{align}
In the second equality we used \eqref{eq:MollerMorphismIntertwiner} and in third equality \eqref{eq:Field}.

As we argued above, $\phi_{\bar A}(g) \doteq V_{\bar A}(\bar A^g - \bar A)$ is a 
c-number for a compactly supported gauge transformation $g$. A natural question is 
now whether this c-number only depends on $g$, or also on the background $\bar A$. 
Using the fact that the implementer is a field, \eqref{eq:V_Field}, 
it is straightforward to show that indeed
\beq
\label{eq:GaugeImplementer}
 \phi_{\bar A}(g) = \phi_{\bar B}(g)
\eeq
if $\bar B - \bar A$ is compactly supported. We use \eqref{eq:V_Consistency} 
with $A = \bar A^g - \bar A$, $\bar B = \bar A + A'$ and $\bar B^g = \bar A + A''$, obtaining
\[
 \tau^\ret_{\bar A, \bar A^g}(V_{\bar A^g}(\bar B^g - \bar A^g)) \phi_{\bar A}(g) = \phi_{\bar B}(g) V_{\bar A}(\bar B - \bar A).
\]
The equality \eqref{eq:GaugeImplementer} then follows from \eqref{eq:MollerMorphism} and \eqref{eq:V_Field}.

\begin{remark}
For space-times $M$ with non-trivial topologies, the possible obstructions 
for the absence of global anomalies are different.
Technically, these obstructions are classified by
the quotient set $\Gamma^\infty_c(M, P\times_{\mathrm{Ad}} G)/\!\!\sim$,
where $P\times_{\mathrm{Ad}} G$ is the Lie group bundle associated to the
adjoint action of the structure group $G$ on itself and $\sim$ is the equivalence relation
given by compactly supported homotopy equivalences.
In the case of $M\simeq \mathbb{R}^d$, this reduces to the $d$-th
homotopy group $\pi_d(G)$. Let us consider now for example $M \simeq \mathbb{R}^{d-1}\times \mathbb{S}^1$
with the trivial principal $G$-bundle $P\simeq M \times G \to M$. The obstructions
for the absence of global anomalies are then given by
$\Gamma^\infty_c(M, P\times_{\mathrm{Ad}} G)/\!\!\sim 
\,\,\, \simeq \, C_c^\infty(\mathbb{R}^{d-1}\times \mathbb{S}^1,G)/\!\!\sim$,
which contains the $d{-}1$-th homotopy group $\pi_{d-1}(G)$ as a subset.
In fact, any element of $\pi_{d-1}(G)$  can be represented (by definition) 
by a compactly supported function $f : \mathbb{R}^{d-1}\to G$, which we can
extend to a compactly supported function $\widetilde{f} : \mathbb{R}^{d-1}\times \mathbb{S}^1\to G$ 
on $M$ that is constant along $\mathbb{S}^1$. This defines an element 
$[\widetilde{f}]\in  C_c^\infty(\mathbb{R}^{d-1}\times \mathbb{S}^1,G)/\!\!\sim$.
It is easy to see that for $f_1,f_2 : \mathbb{R}^{d-1}\to G$ representing different homotopy classes,
the classes $[\widetilde{f_1}]\neq [\widetilde{f_2}]$ in $C_c^\infty(\mathbb{R}^{d-1}\times \mathbb{S}^1,G)/\!\!\sim$
are different: Any compactly supported homotopy $H : [0,1]\times \mathbb{R}^{d-1}\times \mathbb{S}^1\to G$
between $\widetilde{f_1}$ and $\widetilde{f_2}$ induces a compactly supported homotopy 
$H_p :  [0,1]\times \mathbb{R}^{d-1}\to G$ between $f_1$ and $f_2$ by 
fixing any point $p\in\mathbb{S}^1$ of the circle. The same argument generalizes to space-times
of the form $M \simeq \mathbb{R}^{d-k}\times K$, with $K$ a compact $k$-dimensional manifold,
equipped with the trivial principal $G$-bundle. A subset of the obstruction for the absence of global anomalies 
in this case is then given by the $d{-}k$-th homotopy group $\pi_{d-k}(G)$.
\end{remark}

The calculation of the anomaly amounts to the calculation 
of the implementers $V$. In the following, we give a formal expressions for these, whose convergence is unclear. Except for Section~\ref{sec:1d}, where convergence is indeed guaranteed, these expressions will not further be used.
For free fermions, the implementers are formally given by the inverse of the formal $S$ matrix, i.e.,
\begin{equation}
\label{eq:Implementer}
 V_{\bar A}(A) = \TO(e^{\frac{i}{\hbar} j_{\bar A}(A)})^{-1},
\end{equation}
where $\TO$ denotes the time-ordered product, \cf \cite{HollandsWaldStress}.
To see this, we differentiate:
\begin{align*}
 \frac{\ud}{\ud \lambda} V_{\bar A}(\lambda A) 
 & = - V_{\bar A}(\lambda A)\, \frac{\ud}{\ud \lambda} \TO(e^{\frac{i}{\hbar} j_{\bar A}(\lambda A)}) \, V_{\bar A}(\lambda A) \\
 & = - \frac{i}{\hbar} V_{\bar A}(\lambda A)\,  \TO( j_{\bar A}(A) \, e^{\frac{i}{\hbar} j_{\bar A}(\lambda A)}) \, V_{\bar A}(\lambda A) \\
 & = - \frac{i}{\hbar} \Ret(j_{\bar A}(A); e^{\frac{i}{\hbar} j_{\bar A}(\lambda A)})\, V_{\bar A}(\lambda A) \\
 & = - \frac{i}{\hbar} \tau^\ret_{\bar A, \bar A + \lambda A} (j_{\bar A + \lambda A}(A))\,  V_{\bar A}(\lambda A).
\end{align*}
In the third line, the retarded product defined by
\[
 \Ret(F; e^{i S}) \doteq \TO(e^{i S})^{-1} \,\TO(F \otimes e^{i S})
\]
was introduced. In the last step, we used the integral form of perturbative agreement, 
\cf \cite{BackgroundIndependence}, and the fact that the current does not depend 
on the background connection, \cf \eqref{eq:FermionCurrent}. Hence, this implementer fulfills the initial condition 
\eqref{eq:V_0} and the differential equation \eqref{eq:delV}. However, it should be kept 
in mind that it is not clear whether the representation \eqref{eq:Implementer} converges 
(it is derived from a formal power series, but no longer has a formal parameter).

Leaving the convergence question aside for a moment, one could thus calculate the 
implementer by calculating the connected fermion loops with $n$ external legs. 
For simplicity, one may choose $\bar A$ as a trivial connection. From the fermion loops, 
only the non-local part may contribute, as the local terms are gauge invariants, 
and the field strength corresponding to $A$ vanishes.
For the c-number part of the implementer, which is the relevant one for 
the global anomaly, one then has
\begin{align*}
 \bra{\Omega} V_{0}(A)^{-1} \ket{\Omega} & = \sum_{n=0}^\infty \frac{(i/\hbar)^n}{n!} \bra{\Omega} \TO( j(A)^{\otimes n}) \ket{\Omega} \\
 & = \sum_{n=0}^\infty \frac{i^n}{n!} \sum_{2 k_2  + \dots + n k_n = n; k_i \geq 0} \frac{n!}{k_2! \dots k_n!} \prod_{j=2}^n \frac{1}{j^{k_j}} I_j(A)^{k_j} \\
 & = \exp \left( \sum_{k=2}^\infty \frac{i^k}{k} I_k(A) \right),
\end{align*}
where $I_n(A)$ is the closed fermion loop Feynman diagram with $n$ external gauge boson lines, 
each smeared with $A$. 

\begin{remark}
Analogously, one can define global gravitational anomalies 
\cite{AlvarezGaumeWitten}, by replacing changes in the background connection by 
changes in the background metric, gauge transformations by diffeomorphisms, 
and the current by the stress-energy tensor. 
\end{remark}

\section{Hilbert space representations}
\label{sec:Representations}

Let us discuss some consequences of our definitions for representations of our algebras 
on Hilbert spaces. Assume we have a representation $\bar \pi : \A(\bar A) \to \End(\bar \HS)$. 
Consider now arbitrary compactly supported perturbations $A$ of $\bar A$. The 
representation $\bar \pi$ naturally induces representations
\[
 \pi_A \doteq \bar \pi \circ \tau^\ret_{\bar A, \bar A + A} : \A(\bar A + A) \longrightarrow \End(\HS_A),
\]
where $\HS_A = \bar \HS$. Physically, this means that we identify states that coincide in the past. 
In particular, using the canonical identification of $\HS_A$ and $\HS_{A'}$, the algebra 
homomorphism $\tau^\ret_{\bar A + A', \bar A + A}$ is implemented by the identity, i.e.\
\[
 \pi_{A'} \circ \tau^\ret_{\bar A + A', \bar A + A} = \pi_A.
\]
However, there is also another natural map between the algebras 
$\A(\bar A + A)$ and $\A(\bar A + A')$, namely the advanced M{\o}ller operator. 
Let us see whether this can also be implemented:
\begin{align*}
 \pi_{A'} \circ \tau^\adv_{\bar A + A', \bar A + A} & = \pi_{A} \circ \tau^\ret_{\bar A + A, \bar A + A'} \circ \tau^\adv_{\bar A + A', \bar A + A} \\
 & = \pi_A \circ \rce_{\bar A + A}(A'-A) \\
 & = \pi_A \circ \Ad_{V_{\bar A + A}(A'-A)}.
\end{align*}
Hence,
\[
 \pi_{A'} \circ \tau^\adv_{\bar A + A', \bar A + A}(\,\cdot\,) = U(A', A)\, \pi_A( \,\cdot\, )\, U(A', A)^*
\]
with
\begin{align}
 U(A', A) & = \pi_A(V_{\bar A+A}(A'-A)) \nn\\
 & = \bar \pi \left( \tau^\ret_{\bar A, \bar A + A} (V_{\bar A+A}(A'-A)) \right) \nn \\
 & = \bar \pi (V_{\bar A}(A') V_{\bar A}(A)^*), \label{eq:Def_U}
\end{align}
where in the last step we used \eqref{eq:Vcondition}. In particular, we obtain
\begin{align*}
 U(A'', A') U(A', A) & = U(A'', A), \\
 U(A', A)^{-1} & = U(A, A').
\end{align*}
Let us note that these natural properties are a direct consequence of \eqref{eq:Vcondition}, 
which lends further support to the usefulness of this condition.

From \eqref{eq:GaugeImplementer}, it is clear that for a compactly supported gauge transformation $g$
and $A^g = \Ad_{g^{-1}}  A$
\[
\rho(g) \doteq U(\bar A^g + A^g - \bar A, A)
\]
is a c-number independent of the choice of a compactly supported $A$. 
In the case of a global anomaly, there are $g$'s such that $\rho(g) \neq 1$. 
How can these turn a theory inconsistent? There is certainly no problem as
long as the gauge fields are considered purely as background fields. As in the 
path integral approach, one expects inconsistencies in the non-perturbative 
interacting theory. Let us sketch how these manifest themselves. Of course, due to 
the absence of well-developed non-perturbative techniques for quantum field theory, some
of our arguments below necessarily have to be rather formal and heuristic.

Let us consider the interacting Lagrangean 
$\frac{1}{2 g_\YM^2} F \wedge * F + \bar \psi \Dslash \psi$
and split the gauge field  as $\bar A + g_\YM\, A$, where $\bar A$ is a fixed background and 
$A$ is a perturbation. We assume that we can quantize the system consisting of $A$ and $\psi$,
which leads to an algebra of observables $\A(\bar A)^{\mathrm{tot}}$ and a
representation $\bar \pi$ on a Hilbert space $\bar\HS^{\mathrm{tot}}$.
In the free limit $g_{\YM}\to 0$, the gauge field perturbation $A$ decouples from $\psi$,
hence the algebra of observables factorizes  as $\A(\bar A)^{\mathrm{tot}} = 
\A(\bar A)^{\mathrm{gauge}}\otimes \A(\bar A)^\mathrm{matter}$ into gauge field and matter observables,
and the Hilbert space factorizes as $\bar \HS^\mathrm{tot} = 
\bar\HS^\mathrm{gauge}\otimes \bar \HS^{\mathrm{matter}}$. (Notice that the matter algebra of observables
and Hilbert space are the ones we studied above.)

The interpretation of gauge transformations changes in the present setup: While before gauge transformations
were unitary maps $\HS_A \to \HS_{A'}$ between the Hilbert spaces for the matter theory in different
background fields, the fact that we treat gauge fields dynamically requires that both $A$ and $\psi$
have to be transformed. Hence, gauge transformations in the present setup
are unitary maps $\bar \HS^{\mathrm{tot}} \to \bar \HS^{\mathrm{tot}}$
from the Hilbert space of the total theory to itself. In the free limit $g_{\YM}\to 0$, these unitary maps
are supposed to factorize as
\[
 \rho^\mathrm{tot}(g) = \rho^{\mathrm{gauge}}(g)\otimes \rho^{\mathrm{matter}}(g) : 
\bar\HS^\mathrm{gauge}\otimes \bar \HS^{\mathrm{matter}}\to \bar\HS^\mathrm{gauge}\otimes \bar \HS^{\mathrm{matter}},
\]
where for the matter theory we have from above
\beq
\label{eq:rho_g}
 \rho^{\mathrm{matter}}(g) = U(g^{-1} \ud g, 0).
\eeq
For a gauge transformation $g$ that is connected to the identity, this yields the identity operator. 
However, in the case of a global anomaly, this is $e^{i \phi}\, \id$ for a non-trivial phase $\phi$. 
For the representation $\rho^{\mathrm{gauge}}(g)$ on the free gauge fields, 
one expects that no global anomalies are present, 
i.e., gauge transformations always act as the identity on the corresponding Hilbert space, 
so that $\rho^\mathrm{tot}(g) = \id\otimes U(g^{-1} \ud g, 0)$ by \eqref{eq:rho_g}. Switching 
on the interaction, and assuming continuity in $g_\YM$, the phase factor $e^{i \phi}$ must remain constant, 
as it can only take discrete values. Hence, in the case of a global anomaly, 
we find that there is no gauge invariant vector in the total Hilbert space $\bar \HS^{\mathrm{tot}}$.

The almost trivial character of the representation $\rho^\mathrm{tot}$ may lead to the 
suspicion that $\rho^\mathrm{tot}$ is simply not the correct representation to consider. 
After all, why not take the trivial representation? To motivate this further, consider, 
following \cite{Witten82}, the situation in a canonical formalism, i.e., consider a fixed 
time-slice $\Sigma$, defined, for instance, as $x^0 = 0$, and the ``Hilbert space'' of $L^2$ 
functions on the space of initial data on $\Sigma$. Gauge transformations are then generated by the operators
\[
 Q^a(\vec x) = g_\YM^{-2} D_i F^a_{0i}(\vec x) - \bar \psi \gamma^0 T^a \psi(\vec x),
\]
which are to be smeared with compactly supported Lie algebra valued 
functions $\Lambda(\vec x)$. Gauge invariance then dictates that this has to vanish on all physical states.

Let us again restrict to the free limit and the contribution of the matter part. 
Assuming that $j_{\bar A}$ can also be smeared with test ``functions'' localized on a time-slice,
the matter contribution to $Q(\Lambda)$ is $j_{\bar A}(B)$ with
\beq
\label{eq:TimeSliceA}
 B^a_\mu(x) = \delta^0_\mu \Lambda^a(\vec x) \delta(x^0).
\eeq
It easily follows from \eqref{eq:E=0} that this indeed yields a representation of the group $\mathcal{G} = C^\infty_c(\Sigma, G)$ of spatial gauge transformations. The action of the retarded M{\o}ller operator on $\psi$ is given by
\[
 \tau^\ret_{\bar A + B, \bar A} \psi(x) = \psi(x) - \theta(x^0) \Lambda(\vec x) \psi(x) + \dots,
\]
where $\theta$ is the step function and the dots stand for terms that vanish as $x^0 \to 0$. Taking the time-slice which is used to define $Q(\Lambda)$ at $\eps$, one thus obtains, using \eqref{eq:E=0} and the explicit form \eqref{eq:FermionCurrent} of the current,
\[
 [Q(\Lambda), Q(\Lambda')] = i \hbar Q([\Lambda, \Lambda'])
\]
 in the limit $\eps \to 0$.\footnote{The order is irrelevant here, i.e., one could approach 
 $\eps = 0$ either from positive or negative $\eps$, due to the first two terms on the \rhs of \eqref{eq:E=0}.} 
 This may be seen as yet another indication for the appropriateness of condition \eqref{eq:Vcondition}.
 For $\Sigma$ homeomorphic to $\R^{d-1}$, we have $\pi_1(\mathcal{G}) = \pi_d(G)$, 
 so non-trivial space-time gauge transformations correspond to non-trivial cycles of spatial gauge transformations. 
 Integrating up the generators $Q$, i.e., $j$, along such a cycle yields not the identity, but again $e^{i \phi} \, \id$ 
 in the case of a global anomaly.\footnote{Concretely, consider the non-trivial gauge transformation $g$ in $\pi_4(G)$, and the corresponding interpolator $A_\lambda = \lambda g^{-1} \ud g$ in a trivial background $\bar A = 0$. Squeezing $g$ to be supported on the time-slice $x^0 = 0$, for example by defining $g_\eta(x^0, \vec x) = g(x^0 / \eta, \vec x)$ and taking the limit $\eta \to 0$, leads to $A_\lambda$ of the form \eqref{eq:TimeSliceA}. Integrating up the generators, yields, by \eqref{eq:POIntegral}, the non-trivial implementer of the non-trivial gauge transformation.}
 Once more, we conclude that there are no gauge invariant vectors in the Hilbert space.

\section{A one-dimensional example}
\label{sec:1d}

As a simple example, where the anomalies can be computed straightforwardly 
and the background changes are properly unitarily implemented, we consider the 
case of real fermions on the line $\mathbb{R}$. This was first studied in \cite{ElitzurEtAl} 
in the setting of Euclidean partition functions. Their results on the classification of 
anomalies are wrong, however, so the present section also serves to correct these.

Choosing a trivialization of the principal $G$-bundle $P\to \mathbb{R}$, or equivalently
fixing the background connection $\bar A=0$, the action for the fermions 
in the background $\bar A + A$ can be written as
\[
 S_{\bar A+A} = \frac{i}{2} \int_{\mathbb{R}} \psi_i \left( \delta_{i j} \del_t + A_{i j} \right) \psi_j \,\ud t~,
\]
where $A_{i j}(t) = \lambda^m_{i j} \, A^m(t)$ (summation over $m$ understood) and $\lambda^m$ 
are the real antisymmetric generators of the Lie algebra in a representation $r$. 
In the background $\bar A=0$, the $\psi$'s satisfy the equation of motion
$i\,\partial_t \psi=0$ and fulfill the anti-commutation relations 
$\{ \psi_i(t), \psi_j(t^\prime) \} = \hbar\,\delta_{ij}$.
The corresponding current is
\[
 j(A') = \frac{i}{2} \int_\mathbb{R} \psi_i \, A'_{i j}\,  \psi_j \,\ud t.
\]
As there are no ordering ambiguities, this current is conserved, and in particular 
condition \eqref{eq:E=0} is fulfilled. Moreover, the implementers $V(A)$ 
are given by, \cf \eqref{eq:Implementer},
\[
 V(A) = \bar \TO \exp \left( \frac{1}{2\hbar} \int_{\mathbb{R}} \psi_i \,A_{i j} \,\psi_j \,\ud t \right),
\]
where $\bar \TO$ denotes anti time-ordering.

A Hilbert space representation of the algebra of smeared fields
can be obtained by setting
\[
 \bar \pi ( \psi(f) ) \doteq \Psi_i \int_{\mathbb{R}} f_i\, \ud t, 
\]
or, formally, $\bar \pi (\psi_i(t)) = \Psi_i$, where $\Psi_i$ are self-adjoint 
operators on a Hilbert space $\bar \HS$ fulfilling $\{ \Psi_i, \Psi_j \} = \hbar\,\delta_{ij}$.
It follows that the implementers $U(A', A)$ are given by, \cf \eqref{eq:Def_U},
\[
 U(A', A) = \bar \TO \exp \left( \frac{1}{2\hbar} \Psi_i \Psi_j \int_{\mathbb{R}} A'_{i j}\, \ud t \right) 
~ \TO \exp \left( - \frac{1}{2\hbar} \Psi_i \Psi_j \int_{\mathbb{R}} A_{i j}\, \ud t \right).
\]

Following \cite{ElitzurEtAl}, let us now consider the representation 
$R$ on $\bar \HS$ that this induces. For $1 \leq m \leq \rank G$, the generators 
$\lambda$ of the representation $r$ can be brought into the form
\[
 \lambda^m = \begin{pmatrix} 0 & \alpha_1^m & & & \\ - \alpha_1^m & 0 & & & \\ & & 0 & \alpha_2^m &  \\ & & - \alpha_2^m & 0 & \\ & & & &  \ddots \end{pmatrix},
\]
with $\alpha_k$ the weights of $r$. Defining annihilation and creation operators
\begin{align*}
 a_k & \doteq \tfrac{1}{\sqrt{2}} ( \Psi_{2 k - 1} + i \Psi_{2 k} ), &
 a_k^* & \doteq \tfrac{1}{\sqrt{2}} ( \Psi_{2 k - 1} - i \Psi_{2 k} ),
\end{align*}
the generators $\Lambda^m$ in the representation $R$, corresponding 
to the $\lambda^m$ in the representation $r$, are given by
\[
 \Lambda^m = \sum_{k = 1}^{[d_r/2]} \alpha_k^m (a_k^* a_k - \tfrac{1}{2} ),
\]
where $d_r$ is the dimension of the representation $r$ and 
$[ \cdot ]$ denotes the integer part. The weights of $R$ thus have the form
\[
 \delta = \tfrac{1}{2} \sum \pm \alpha_k.
\]
An anomaly occurs if there is a weight of $R$ which is non-integral for the gauge 
group $G$, as then $U(g^{-1} \ud g, 0)$ is not the identity for $g \in \pi_1(G)$. 
A simple example is obtained taking the fermions in the fundamental 
representation of $G = SO(3)$.

The existence of integral weights of $\g$ which are non-integral for 
$G$ is equivalent to a non-trivial $\pi_1(G)$, establishing the relation to the 
abstract result that a global anomaly is possible only for non-trivial 
$\pi_d(G)$, with $d$ being the space-time dimension.\footnote{In \cite{ElitzurEtAl}, 
the occurrence of a global anomaly was erroneously related to the question whether 
the weights of $R$ are in the same equivalence class as those of $r$ modulo the 
root lattice, leading to the conclusion that there are global anomalies for many 
groups with a trivial $\pi_1$.} Of course the example is somewhat artificial, 
as one may remove the anomaly by re-defining the gauge group to 
be the universal cover of $G$.

\section{Perturbative agreement and the Wess-Zumino consistency condition}
\label{sec:WZ}

Let us consider again the condition \eqref{eq:E=0} on the vanishing of the 
curvature $E_{\bar A}$ for $A$, $A'$ being infinitesimal gauge transformations, i.e., 
$A = \bar \ud \Lambda$ and $A' = \bar \ud \Lambda'$. In that case the current 
evaluated at $\bar A$ is a c-number, so that the commutator vanishes. For 
the expression on the \lhs of \eqref{eq:E=0}, we thus obtain
\begin{align}
 \delta^\ret_{\bar A}(\bar \ud \Lambda) j(\bar \ud \Lambda') - \delta^\ret_{\bar A}(\bar \ud \Lambda') j(\bar \ud \Lambda) & = \delta^\ret_{\bar A}(\bar \ud \Lambda) \bar \delta j(\Lambda') - j_{\bar A}([\bar \ud \Lambda, \Lambda']) - \delta^\ret_{\bar A}(\bar \ud \Lambda') \bar \delta j(\Lambda) + j_{\bar A}([\bar \ud \Lambda', \Lambda]) \nn \\
 & = \bar \delta j_{\bar A}([\Lambda, \Lambda']) - \bar \delta j_{\bar A}([\Lambda', \Lambda]) - j_{\bar A}(\bar \ud [\Lambda, \Lambda']) \nn \\
\label{eq:delta_j_Calculation}
 & = \bar \delta j_{\bar A}([\Lambda, \Lambda']),
\end{align}
where the notation introduced in \eqref{eq:CurrentConservation}
was employed. 
In the first step, we used an identity used in the proof of 
\cite[Prop.~3.8]{BackgroundIndependence}. In the second step we used
\beq
\label{eq:GaugeTrafoField}
 \delta^\ret_{\bar A}(\bar \ud \Lambda) \Phi(t) = \Phi_{\bar A}(\Lie_\Lambda t),
\eeq
for any field $\Phi$, where $\Lie_\Lambda$ is the representation of $\Lambda \in \Gamma^\infty_c(M, \p)$ 
on the test tensor. This is a straightforward consequence of \eqref{eq:Field} and \eqref{eq:MollerMorphism}.
From \eqref{eq:delta_j_Calculation}, we see that for semisimple Lie algebras, the vanishing 
of the divergence of the current is a necessary condition for the fulfillment of 
\eqref{eq:E=0}. On the other hand, as shown in \cite{BackgroundIndependence}, the 
vanishing of the divergence of the current implies \eqref{eq:E=0} in space-time dimension $d \leq 4$.

For the computation of the global anomaly in the next section, it is advantageous to embed
the theory into a larger one which has a local anomaly, i.e., a non-vanishing divergence of the current.
However, one still wants to fulfill \eqref{eq:E=0}, in order to be able to deform the integration path in
\eqref{eq:POIntegral}. We thus investigate whether we can
save \eqref{eq:E=0} by giving up the requirement that the current is a field, i.e., 
by giving up covariance \wrt transformations of the background field. This means that we fix 
a reference connection and specify other connections by a Lie algebra valued one form $A$. 
For simplicity, we assume the reference connection to be flat. The current $j_A$ is then allowed 
to depend on $A$, i.e., we take the covariant current $j$
and add an $A$ dependent correction term $\Delta j_A$. In the 
calculation \eqref{eq:delta_j_Calculation}, we can then no longer use \eqref{eq:GaugeTrafoField}. 
Instead, in the expression on the \rhs of the first line of \eqref{eq:delta_j_Calculation}, we may 
use that $\bar \delta j$ is a c-number, so that the retarded variation coincides with the 
functional derivative \wrt $A$. Hence, we obtain the condition
\beq
\label{eq:WZ}
 \skal{\tfrac{\delta}{\delta A} (\delta j)_A(\Lambda')}{\ud_A \Lambda} - \skal{\tfrac{\delta}{\delta A} (\delta j)_A(\Lambda)}{\ud_A \Lambda'} - (\delta j)_{A}([\Lambda, \Lambda']) = 0,
\eeq
where $\ud_A$ is the covariant differential in the background $A$ and $(\delta j)_A(\Lambda) \doteq j_A(\ud_A \Lambda)$.
This is the \emph{Wess-Zumino consistency condition} \cite{WessZumino}, derived from the 
requirement that gauge transformations of the background field are represented on 
the vacuum functional. Hence, after giving up background gauge invariance, i.e., the 
requirement that physics should only depend on the principal bundle connection, not on 
a particular choice of a trivialization, it may be possible to get rid of the path dependence 
in the definition of the implementers, even in the presence of local anomalies.

In \cite{BackgroundIndependence}, it was shown that in dimension $d \leq 4$, the vanishing 
of the divergence of the current implies that \eqref{eq:E=0} is fulfilled also for variations 
$A$, $A'$ which are not gauge. We will prove an analogous statement for the modified 
current, i.e., that \eqref{eq:WZ} implies \eqref{eq:E=0} for non-gauge variations 
$A$, $A'$. However, we shall explicitly use the assumption that the gauge group is semisimple 
and that the (covariant) anomaly is of the form
\beq
\label{eq:CovariantAnomaly}
 \bar \delta j_{\bar A}(\Lambda) = c \int_M \tr \big(\Lambda \bar F \wedge \bar F \big),  
\eeq
where the trace is taken in some representation. 
As a first step, we claim that
\beq
\label{eq:E_A_A'}
 E_{\bar A}(A, A') = c \int_M \tr\big( (A \wedge A' - A' \wedge A) \wedge \bar F \big)
\eeq
as this is the unique functional linear and antisymmetric in $A$, $A'$, invariant under 
$\delta \bar A = \bar \ud \Lambda$, $\delta A^{(\prime)} = [\Lambda, A^{(\prime)}]$ 
and of the correct scaling dimension, such that
\begin{align*}
 E_{\bar A}(\bar \ud \Lambda, \bar \ud \Lambda') & = c \int_M \tr \big( ( \bar \ud \Lambda \wedge \bar \ud \Lambda' - \bar \ud \Lambda' \wedge \bar \ud \Lambda) \wedge \bar F \big)\\
 & = c \int_M \tr\big( [\Lambda, \Lambda'] \bar F \wedge \bar F \big)\\
 & = \bar \delta j_{\bar A}([\Lambda, \Lambda']).
\end{align*}
Uniqueness follows from the fact that $E_{\bar A}(\bar \ud \Lambda, \bar \ud \Lambda') = 0$ 
implies $E_{\bar A}(A, A') = 0$ as proved in \cite{BackgroundIndependence}.

It is well-known \cite{BardeenZumino} how to pass from the covariant anomaly 
\eqref{eq:CovariantAnomaly} to the consistent anomaly fulfilling the Wess-Zumino 
consistency condition \eqref{eq:WZ}. One adds to the current the correction term
\beq
\label{eq:Delta_j}
 \Delta j_A(A') \doteq \tfrac{1}{3} c \int_M \tr \big( A' \wedge ( A \wedge F_A + F_A \wedge A - \tfrac{1}{2} A \wedge A \wedge A ) \big).
\eeq
For the total anomaly, we thus obtain
\begin{align*}
 (\delta j)_A(\Lambda) & = c \int_M \tr \left( \Lambda F_A \wedge F_A + \tfrac{1}{3} \ud_A \Lambda \wedge (A \wedge F_A + F_A \wedge A - \tfrac{1}{2} A \wedge A \wedge A) \right) \\
  & = \tfrac{1}{3} c \int_M \tr \left( \Lambda (  \ud A \wedge \ud A + \tfrac{1}{2} \ud (A \wedge A \wedge A) ) \right),
\end{align*}
which is the well-known expression for the consistent anomaly. Adding \eqref{eq:Delta_j} modifies the curvature $E$ by
\begin{align*}
 \Delta E_A(A_1, A_2) & = \skal{\tfrac{\delta}{\delta A} \Delta j_A(A_2)}{A_1} - \skal{\tfrac{\delta}{\delta A} \Delta j_A(A_1)}{A_2} \\
  & = - c \int_M \tr \left( (A_1 \wedge A_2 - A_2 \wedge A_1) \wedge F_A \right).
\end{align*}
Comparison with \eqref{eq:E_A_A'} shows that the curvature $E$ is indeed canceled.

\begin{remark}
In \cite{HollandsWaldStress}, the violation of perturbative agreement was related to a 
cohomological question by noting that $E$ is a cocycle \wrt a suitably defined differential. 
It was shown that a violation of perturbative agreement can be removed if $E$ is a coboundary. 
At least for the Yang-Mills case, we have seen that $E$ is always a coboundary, but in general 
of a background field functional depending on the choice of a trivialization of the background.
\end{remark}

\begin{remark}
There is a subtle point here: For $G = U(1)$ and on flat space-time, one could 
remove the anomaly by replacing the factor $\frac{1}{3}$ in \eqref{eq:Delta_j} 
by $\frac{1}{2}$. In this way, one also fulfills the Wess-Zumino consistency 
condition \eqref{eq:WZ}. However, the curvature $E$ does then not vanish 
(except when both arguments are exact). This shows that the condition 
\eqref{eq:E=0} is stronger than the Wess-Zumino consistency condition 
\eqref{eq:WZ} for non semisimple $G$ and non-generic space-times.
\end{remark}

\section{Computation of the anomaly}
\label{sec:Computation}

For the actual computation of the anomaly, we follow the ideas of \cite{Witten83a, ElitzurNair}:
On a topologically trivial $n$-dimensional spacetime, this method requires that one may embed
$G$ as a subgroup in $H$, with trivial $\pi_n(H)$. Furthermore, there must exist a representation
of $H$ which upon reduction to $G$ yields the original representation up to the addition of
trivial representations. Under these conditions, the global anomaly of $G$ may be computed by
integrating up the local anomaly of $H$.

Let us consider the example of $G = SU(2)$ in the fundamental representation in $n = 4$.
We may embed $G$ as a subgroup in $H = SU(3)$ and extend the fundamental
representation of $G$ to the one of $H$, such that a $G$-gauge transformation $g$ is given by
\[
 g_H = \begin{pmatrix} g & 0 \\ 0 & 1 \end{pmatrix}.
\]
As $\pi_4(H)$ is trivial, we may connect the non-trivial $G$-gauge transformation
$g$ (i.e.\ $[g]=-1\in\pi_4(G)\simeq \mathbb{Z}_2$)
by a path $h(\lambda)$ of $H$-gauge transformations to the identity, i.e.,
\begin{equation}\label{eqn:gaugepathH}
 h(0) = \id, \qquad h(1) = g_H.
\end{equation}
The global anomaly of $G$ can then be computed by integrating up the 
local anomaly of $H$ along this path, using \eqref{eq:POIntegral}. For this to work, 
two requirements have to be fulfilled: First, it must be possible to deform the path 
$h$ to a path with background connections solely in the $G$ component, i.e., of the form
\beq
\label{eq:A_H}
 A_H = \begin{pmatrix} A & 0 \\ 0 & 0 \end{pmatrix}.
\eeq
The path-ordered exponential \eqref{eq:POIntegral} is path independent if \eqref{eq:E=0}
is fulfilled. However, the fundamental representation of $SU(3)$ has a local anomaly, given by
\[
 \bar \delta j(\Lambda) = \frac{i \hbar}{8 \pi^2} \int_M \tr \big(\Lambda \bar F \wedge \bar F\big)
\]
for a single generation of chiral fermions,\footnote{Note that this can be computed in the Lorentzian setting 
\cite{ChiralFermions}, without recourse to Riemannian concepts.} so that \eqref{eq:E=0} is not fulfilled.
According to the discussion
in the previous section, we may nevertheless achieve \eqref{eq:E=0} through a redefinition of the
current (breaking local gauge covariance), yielding
a current with the consistent anomaly
\begin{equation}\label{eqn:consistentanomalyexplicit}
 (\delta j)_A(\Lambda) = \frac{i \hbar}{24 \pi^2} \int_M \tr \left( \Lambda \, \ud ( A \wedge \ud A + \tfrac{1}{2} A \wedge A \wedge A ) \right).
\end{equation}
It is thus this current which has to be integrated in \eqref{eq:POIntegral}.
The second requirement is that the implementer for perturbations 
of the form \eqref{eq:A_H} must reduce to the implementer in the original (unextended) theory. 
In particular, local gauge covariance must be restored, i.e., it must be independent of the choice 
of a trivialization.

Let us sketch such a construction:\footnote{This paragraph is rather technical and 
deals with the details of how to construct the algebras $\A$ as evaluation functionals. 
A reader more interested in the structural aspects may safely skip it.}
Given a background connection $\bar A$ of the original 
theory, choose some trivialization, obtain the corresponding $A$ and set the 
background connection of the extended theory to be $A_H$. Given a two-point 
function $\omega$ for the original background $\bar A$, construct a two-point 
function of the extended theory to be
\[
 \omega_H(\Phi, \Psi) = \omega(\phi, \psi) + \omega_0(\Phi_3, \Psi_3),
\]
where $\Phi = (\phi, \Phi_3)$ in the decomposition above, $\Phi$ and $\Psi$ 
are test sections, and $\omega_0$ is some two-point function for the supplementary 
singlet with a vanishing current. We now require that for the implementers 
for changes of the background connection of the form \eqref{eq:A_H}, we have
\beq
\label{eq:ExtensionConsistency}
 V_{\bar A}(A')_{\omega}(\phi) = V_{\bar A_H}(A'_H)_{\omega_H}(\phi, 0).
\eeq
Here we interpreted elements of the algebra as evaluation functionals 
(with $\phi$ a configuration) and used the ``quantum functional'' notation 
of \cite{LocCovDirac}, where in order to make local covariance explicit,
one does not work with an algebra with a fixed $\star$ product derived 
from a two-point function $\omega$, but with flat sections over the set 
of all Hadamard two-point functions (flatness being defined by the canonical 
equivalence relation between different Hadamard two-point functions). 
The way to prove \eqref{eq:ExtensionConsistency} is by noting that 
the implementers on both sides obey the same ODE if one reaches 
$A'$ by a path of background changes. Namely, by assumption, 
$\omega_0$ does not contribute to the current and also the correction 
term \eqref{eq:Delta_j} does not contribute, as the original theory 
had no local anomalies (the constants $d_{abc}$ vanish).

To compute the global anomaly on a trivial background connection, we proceed as follows:
Given any path $h(\lambda)$ of $H$-gauge transformations as in \eqref{eqn:gaugepathH},
we set
\beq
\label{eq:A_h}
 A = h^{-1} \ud h,
\eeq
where $\ud$ denotes the four-dimensional differential. Its derivative \wrt $\lambda$ is given by
\beq
\label{eq:A'_h}
 \dot A = \del_\lambda h^{-1} \ud h + h^{-1} \ud \del_\lambda h = \ud \Lambda + [A, \Lambda]
\eeq
with
\beq
\label{eq:Lambda_h}
 \Lambda = h^{-1} \del_\lambda h.
\eeq
Integrating the consistent anomaly \eqref{eqn:consistentanomalyexplicit} by using \eqref{eq:POIntegral}, 
we obtain the following candidate for the global $SU(2)$ anomaly
\begin{flalign}
\nn 
 V_0(g_H^{-1}\ud g_{H}) &= \exp \left( \frac{1}{48 \pi^2} \int_0^1 \ud \lambda \int_M \tr \left(\Lambda\,A\wedge A\wedge A\wedge A\right)  \right) \\
 &= \exp \left( \frac{1}{240 \pi^2} \int_{M\times[0,1]} h^\ast\big( \mu_H^5\big)\right),\label{eq:WittenAnomaly}
\end{flalign}
where we used that the connection $A$ has vanishing curvature, i.e.\ $\ud A = - A \wedge A$. 
Moreover, 
\[
\mu_H^5 \doteq \tr \big(\mu_H\wedge\mu_H\wedge\mu_H\wedge\mu_H\wedge\mu_H\big)\in\Omega^5(H)
\]
is the Cartan $5$-form on $H=SU(3)$ and $h^\ast(\mu_H^5)\in \Omega^5(M\times[0,1])$ denotes
its pullback along our path of gauge transformations $h : M\times[0,1] \to H$.

It remains to show that \eqref{eq:WittenAnomaly} implies that $V_0(g_H^{-1}\ud g_{H}) = -\oone$, 
i.e.\ that the global $SU(2)$ anomaly may be detected in our approach. 
Recalling that our path of gauge transformations $h : M\times[0,1] \to H$ 
satisfies the boundary conditions in \eqref{eqn:gaugepathH}, it defines an element in
the $5$-th homotopy group $\pi_5(H/G)$ of the quotient $H/G = SU(3)/SU(2)$.
The integral in \eqref{eq:WittenAnomaly} does not depend on the choice of representative, hence
it defines a mapping
\begin{equation}\label{eqn:grouphom}
\pi_5(H/G) \longrightarrow \mathbb{R}~,~~[h] \longmapsto \frac{1}{240 \pi^2} 
\int_{\mathbb{S}^5} h^\ast\big( \mu_H^5\big),
\end{equation}
which is easily seen to be a group homomorphism.
Using as in \cite{ElitzurNair} the following exact sequence of homotopy groups
\begin{equation}\label{eqn:sequence}
\xymatrix{
\pi_5(H) \ar[r]& \pi_5(H/G) \ar[r] & \pi_4(G) \ar[r] & \pi_4(H)
}~,
\end{equation}
together with the normalization
\begin{equation}\label{eqn:normalization}
\frac{1}{240 \pi^2}  \int_{\mathbb{S}^5} h_1^\ast\big( \mu_H^5\big) = 2\pi\,i
\end{equation}
for the generator $[h_1]$ of  $\pi_5(H)\simeq \mathbb{Z}$,
we obtain the desired result that $V_0(g_H^{-1}\ud g_{H}) = -\oone$.
Let us explain this in more detail: 
Our non-trivial large $G=SU(2)$-gauge transformation $g$ represents by definition
the non-trivial homotopy class $[g]=-1\in \pi_4(G)\simeq \mathbb{Z}_2$. 
Using the exact sequence \eqref{eqn:sequence} and $\pi_4(H) =0$ for $H=SU(3)$,
we obtain a preimage of this class in $\pi_5(H/G)$,
which we may represent by a path $h$ of $H$-gauge transformations as in \eqref{eqn:gaugepathH}.
As the corresponding homotopy class $[h]\in \pi_5(H/G)\simeq \mathbb{Z}$ is an {\em odd} number,
the group homomorphism property of \eqref{eqn:grouphom} 
together with the normalization \eqref{eqn:normalization} implies that
the exponent in \eqref{eq:WittenAnomaly} is an {\em odd} multiple of $\pi\,i$.
Hence, $V_0(g_H^{-1}\ud g_{H}) = -\oone$ and we have detected the global $SU(2)$ anomaly.
\begin{remark}
As discussed in Section~\ref{sec:GlobalAnomalies}, the global anomaly is not changed by 
compactly supported modifications of the background connection. Hence, it is conceivable 
that it depends on the asymptotic behavior of the background connection. This, however, 
is not the case for the global $SU(2)$ anomaly, i.e.,
for a generic background connection $\bar A$ we have
$V_{\bar A}(\bar A^{g} -\bar A) = -\oone$, where 
$g$ is a $G$-gauge transformation representing the non-trivial homotopy class
$[g] =-1\in \pi_4(G)\simeq \mathbb{Z}_2$.
Instead of \eqref{eq:A_h}, in the present situation we have to take 
\[
A = \bar A^h -\bar A,
\]
where $h$ is a path of $H$-gauge transformations as in \eqref{eqn:gaugepathH}.
Instead of \eqref{eq:A'_h}, the $\lambda$-derivative then reads as
\[
\dot A = \ud \Lambda + [\bar A^h,\Lambda]
\]
with $\Lambda$ given by \eqref{eq:Lambda_h}.
It is convenient to regard $\Lambda$ as the $5$-th component of
the $5$-dimensional $H$-gauge potential on $M\times [0,1]$
given by
\begin{equation}\label{eqn:5gaugepot}
\mathbf{A}^h \doteq \bar A^h +\Lambda\,\ud \lambda = h^{-1} \bar A h + h^{-1}\mathbf{d} h,
\end{equation}
where $\mathbf{d}$ is the $5$-dimensional de Rham differential on $M\times [0,1]$.
Integrating the consistent anomaly \eqref{eqn:consistentanomalyexplicit} 
by using \eqref{eq:POIntegral}, we obtain after some simplifications
\[
V_{\bar A}(\bar A^{g}-\bar A) = \exp\left(\frac{1}{24\pi^2} \int_{M\times [0,1]} \mathrm{CS}_5(\mathbf{A}^h)\right),
\]
where
\[
\mathrm{CS}_5(\mathbf{A}^h)=\tr \left(\mathbf{F}^h\wedge\mathbf{F}^h \wedge \mathbf{A}^h  -\frac{1}{2}\, \mathbf{F}^h\wedge \mathbf{A}^h\wedge\mathbf{A}^h\wedge \mathbf{A}^h 
 + \frac{1}{10}
\mathbf{A}^h\wedge\mathbf{A}^h\wedge\mathbf{A}^h\wedge\mathbf{A}^h\wedge\mathbf{A}^h \right)
\]
is the Chern-Simons $5$-form and $\mathbf{F}^h\doteq \mathbf{d} \mathbf{A}^h + \mathbf{A}^h\wedge \mathbf{A}^h$
is the $5$-dimensional curvature. Using that by \eqref{eqn:5gaugepot} $\mathbf{A}^h$ 
is obtained by an $H$-gauge transformation
of the $4$-dimensional gauge potential $\bar A$, the transformation property of Chern-Simons forms
together with $\mathrm{CS}_5(\bar A)=0$ (because $\bar A$ is independent of $\lambda$ and
has no $\ud\lambda$ component) implies that
\[
\frac{1}{24\pi^2} \int_{M\times [0,1]} \mathrm{CS}_5(\mathbf{A}^h) ~=~ \frac{1}{240\pi^2} \int_{M\times [0,1]}  h^\ast(\mu^{5}_H).
\]
Hence, we obtain the same expression for the global $SU(2)$ 
anomaly as in the case $\bar A=0$ above, \cf \eqref{eq:WittenAnomaly}.
\end{remark}

\subsection*{Acknowledgments}
We would like to thank Dirk-Andr\'e Deckert, Chris Fewster, 
Stefan Hollands and Ko Sanders for helpful discussions. 
A.S.~was supported by a Research Fellowship of the 
Deutsche Forschungsgemeinschaft (DFG, Germany).
A large part of the work presented here was done at Heriot-Watt 
University Edinburgh. 
J.Z.~would like to thank the Department of Mathematics 
for the kind hospitality and the COST action 
``Quantum structure of spacetime (QSPACE)'' for funding 
the visit through the ``short term scientific missions'' program.

\end{document}